\documentclass[10pt]{article}
\usepackage[a4paper,margin=22mm]{geometry}
\usepackage[T1]{fontenc}
\usepackage[utf8]{inputenc}
\usepackage{lmodern}
\usepackage{amsmath,amssymb,bm}
\usepackage{graphicx}
\DeclareGraphicsExtensions{.eps}
\graphicspath{{figures/}}
\usepackage{cite}
\usepackage[font=small,labelfont=bf]{caption}
\usepackage[hidelinks]{hyperref}

\title{\textbf{Artificial Anisotropy Induced Bound States in the Continuum for Integrated Photonic Waveguide }}
\author{%
Jinzhao Wang$^{1,\dagger}$, Kunrun Lu$^{1,\dagger}$, Yuanlin Li$^{1}$, Weiming Yao$^{1}$, Yang Feng$^{1}$, Yidi Cao$^{1}$,\\
Wei Liu$^{1}$, Feng He$^{1}$, Jianan Duan$^{1}$, Yi Zou$^{2,*}$, Yongkang Dong$^{1,*}$, Xiaochuan Xu$^{1,3,*}$\\[0.8em]
\small $^{1}$National Key Laboratory of Laser Spatial Information,\\
\small Guangdong Provincial Key Laboratory of Integrated Photonic-Electronic Chip,\\
\small Guangdong Provincial Key Laboratory of Aerospace Communication and Networking Technology,\\
\small Harbin Institute of Technology, Xili University Town, Shenzhen, Guangdong 518055, China\\
\small $^{2}$School of Information Science and Technology, ShanghaiTech University,\\
\small Shanghai 201210, China\\
\small $^{3}$Pengcheng Laboratory, Shenzhen, Guangdong 518055, China\\[0.5em]
\small $^\dagger$These authors contributed equally to this work.\\
\small $^*$Corresponding authors:\\
\small \href{mailto:zouyi@shanghaitech.edu.cn}{zouyi@shanghaitech.edu.cn} (Y.Z.);
\href{mailto:aldendong@163.com}{aldendong@163.com} (Y.D.);\\
\small \href{mailto:xuxiaochuan@hit.edu.cn}{xuxiaochuan@hit.edu.cn} (X.X.)
}
\date{}

\begin{document}
\maketitle

\begin{abstract}
Bound states in the continuum (BICs) enable counterintuitive light confinement without radiation loss, providing a powerful foundation for integrated photonic waveguides. However, existing BIC waveguides are predominantly realized through geometry-dependent designs, where the BIC condition is restricted to narrowly defined structural parameters, limiting design flexibility and practical applicability. Artificial optical anisotropy is introduced as a new design paradigm for BIC waveguides. Implemented using subwavelength-grating (SWG) metamaterials, continuously tailorable anisotropy provides an independent degree of freedom for deterministically reshaping the radiative continuum, enabling flexible formation and systematic control of BIC waveguides over a broad design space. Anisotropy-engineered symmetry breaking further enables controllable asymmetric radiation and precisely tailored field leakage. This paradigm transforms BIC waveguides from geometry-constrained structures into an anisotropy-engineered platform, establishing a general framework for programmable radiation engineering and next-generation integrated photonic devices.
\end{abstract}

\section{Introduction}

Bound states in the continuum (BICs) represent a unique class of waves that remain perfectly confined despite residing within the spectrum of radiating states \cite{ref1,ref2,ref3,ref4}. This counterintuitive phenomenon has established BICs as a powerful mechanism for manipulating light in integrated photonics, enabling applications including low-loss waveguiding \cite{ref5}, ultrasensitive sensing \cite{ref6}, microlasers \cite{ref7}, and polarization manipulation \cite{ref8}. Among these developments, cladding--core--cladding waveguides have emerged as the canonical platform for realizing photonic BICs \cite{ref9,ref10,ref11}. A landmark demonstration was achieved in 2007, when BIC-guided propagation was experimentally observed in silicon ridge waveguides, where transverse-magnetic (TM) modes exhibit vanishing radiation loss at specific core widths through destructive interference of TM--TE mode conversion at the core--cladding interfaces \cite{ref1,ref2,ref12,ref13}. Subsequent studies have exploited this mechanism to realize high-quality resonances \cite{ref5,ref9,ref11,ref14}, high polarization extinction ratios \cite{ref15}, and multimode waveguiding \cite{ref16}.

Despite these advances, the design of BIC waveguides remains fundamentally constrained by a geometry-dependent paradigm. In existing implementations, the BIC condition is achieved only within a narrow range of structural parameters, where geometric tuning simultaneously governs modal birefringence, continuum coupling, and radiation interference \cite{ref17,ref18,ref19}. Consequently, modifying one property inevitably perturbs the others, resulting in a limited design space and restricting the flexibility required for increasingly sophisticated photonic functionalities. Although natural anisotropic materials such as LiNbO\textsubscript{3} provide an additional means of modifying the continuum spectrum, their fixed birefringence offers only limited control \cite{ref20,ref21}, while heterogeneous material integration inevitably increases fabrication complexity \cite{ref5,ref10,ref16}. A fundamentally different strategy capable of independently engineering the radiative continuum is therefore highly desirable.

Artificial optical anisotropy offers such a possibility. Subwavelength gratings (SWGs) behave as homogeneous anisotropic media when operated below the diffraction limit, with effective optical properties accurately described by effective medium theory \cite{ref22,ref23,ref24}. Unlike natural crystals, SWG metamaterials provide continuously tailorable permittivity and birefringence through lithographic patterning, enabling strong and programmable anisotropy within a standard CMOS-compatible fabrication process \cite{ref25}. Moreover, rotating the grating orientation introduces an additional degree of freedom for manipulating the anisotropic optical axis and modal birefringence \cite{ref26,ref27}. These unique capabilities have recently enabled enhanced light confinement and reduced evanescent skin depth \cite{ref28,ref29,ref30}, suggesting that artificial anisotropy may provide a fundamentally new mechanism for controlling BIC waveguides beyond geometric optimization.

Here, artificial optical anisotropy is introduced as a new design paradigm for BIC waveguides by integrating anisotropic SWG metamaterials into the cladding regions of a cladding--core--cladding architecture. Rather than relying on precise geometric tuning, continuously tailorable anisotropy serves as an independent degree of freedom for deterministically reshaping the radiative continuum, enabling flexible formation and systematic control of BIC conditions over a substantially expanded design space. Furthermore, anisotropy-engineered symmetry breaking is realized by rotating the optical axis of the SWG claddings, enabling controllable asymmetric radiation, asymmetric field confinement, and programmable leakage while preserving the interference mechanism underlying BIC formation. This approach transforms BIC waveguides from geometry-constrained structures into an anisotropy-engineered platform, establishing a general framework for programmable radiation engineering and highly flexible integrated photonic devices.

\section{Results and Discussion}

\subsection{Formation of anisotropy-induced BIC waveguide}

Fig.~1A illustrates the cladding--core--cladding waveguide geometry and the corresponding modal effective indices. As the core index is higher than that of the cladding, core-guided modes generally exhibit larger effective indices than cladding slab modes. A critical exception occurs when the effective index of the core TM mode, \(N_{\mathrm{TM}}\), falls below the cladding TE slab mode, \(N_{\mathrm{TE}}^{\mathrm{clad}}\), embedding the TM mode within the radiative continuum formed by the cladding TE slab modes. This results in TM-to-TE conversion at the core--cladding index discontinuity, triggering substantial leakage, as shown in Fig.~1B (see details in Supplementary Section S1). This radiation is not, however, inevitable; net radiation can be suppressed through destructive interference between the TE components under the condition \cite{ref31}:

\begin{equation}W=\frac{\left(m+\Delta\Phi/2\pi\right)\lambda}{\sqrt{\left(N_{\mathrm{TE}}^{\mathrm{core}}\right)^2-N_{\mathrm{TM}}^2}}.\label{eq:bic-condition}\end{equation}

where \emph{m} is a no-zero integer, \(\Delta\Phi\) is the phase difference between reflected and transmitted TE waves, \(\lambda\) is the wavelength, \(N_{\mathrm{TE}}^{\mathrm{core}}\) is the effective index of TE slab mode in the core region. Under this condition, the core TM mode remains fully confined despite being spectrally embedded in the TE-slab continuum, forming a BIC. In conventional implementations, the radiative continuum is established by shallow etching of the cladding slab, offering only the etch depth as a tunable parameter. Such one-parameter control largely predefines the available radiation channels and limits the ability to reshape the leakage pathway, pinning BIC to rigid geometric conditions and impeding application-driven engineering.

Alternatively, the radiative continuum can be engineered laterally by embedding a middle-index SWG slab into the cladding, as schematized in Fig.~1C. Based on the EMT, the SWG slab can effectively behave as a solid-core slab with the equivalent core material described by ordinary and extraordinary refractive indices \(n_{\parallel}\) and \(n_{\perp}\), respectively. These indices can be continuously tuned via the duty cycle \(a/\Lambda\), ranging from air-like to silicon-like regimes (see details in Supplementary Section S2). This engineered anisotropy directly maps onto a tunable modal birefringence, providing a versatile approach to reshape the cladding continuum through polarization-selective dispersion control. The resulting anisotropic response is visualized in Fig.~1D via the normalized wave-vector surfaces of modes with orthogonal polarizations. The dashed curves denote in-plane plane-wave isofrequency contours of a uniaxial crystal, whereas the solid curves represent the guided modes supported by a SWG slab with corresponding effective indices. The TM branch remains circular, indicating that its effective index is largely insensitive to the in-plane propagation direction. In contrast, the TE branch exhibits a pronounced elliptical contour, revealing a strong angular dependence of its effective index. Consistently, the TE mode wave-vector in the SWG slab closely follows the corresponding plane-wave manifold, indicating that the TE slab radiative channel can be efficiently and continuously tuned through anisotropic metamaterial design. This polarization-asymmetric dispersion is central to our strategy: it enables the effective index of the cladding TE slab mode (\(N_{\mathrm{TE}}^{\mathrm{clad}}\)) to be shifted over a wide range by rotating the optical axis, while leaving the TM transmission characteristics comparatively intact. The SWG anisotropy therefore provides a mechanism to reconfigure the radiative continuum and its coupling conditions without compromising the integrity of the radiation channel, establishing a route to deterministic, high-fidelity control of BIC formation.

\begin{figure}[!t]
\centering
\includegraphics[width=\textwidth,height=0.68\textheight,keepaspectratio]{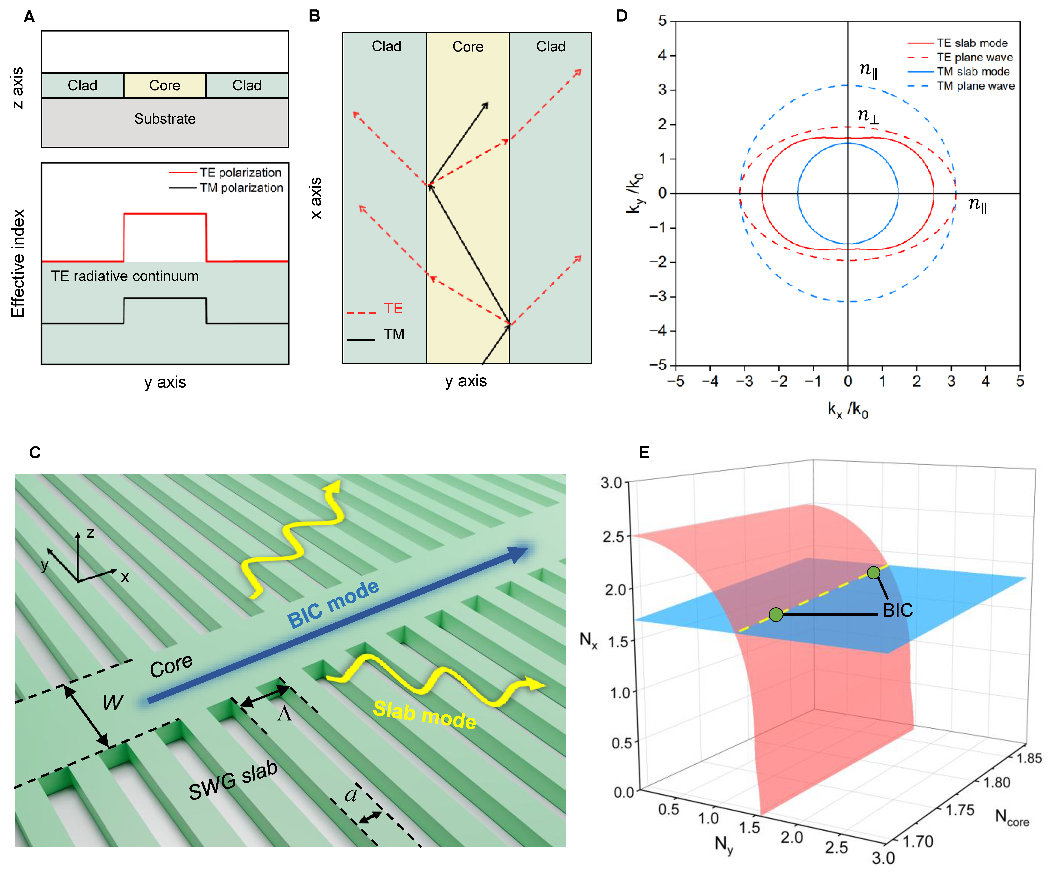}
\caption{\textbf{Concept of the waveguiding enabled by anisotropy-induced BICs.} (\textbf{A)} Cross-section of the cladding-core-cladding waveguide and corresponding modal effective refractive index distribution. The core TM mode lies within the continuum of the cladding TE slab mode. (\textbf{B}) TM-to-TE mode conversion at the core--cladding junction. A transmitted TE wave and a reflected TE wave are generated at each conversion event. (\textbf{C}) Structure of SWG-enabled cladding-core-cladding waveguides where SWGs are integrated into the cladding regions. (\textbf{D}) Normalized wavevectors of the TE/TM modes in the SWG slab and in-plane TE/TM plane waves in the uniaxial crystal with the same anisotropic indices. (\textbf{E}) Illustration of the momentum of modes in artificially anisotropic cladding waveguides, where the red surface represents the cladding TE slab mode and blue surface represents the core TM mode. The yellow dash line corresponds to the phase-match condition where two modes share the same longitudinal component. The discrete green dots represent the emergence of the BIC where the destructive interference of radiative TE waves is satisfied, corresponding to specific effective index of the core TM mode.}
\label{fig:1}
\end{figure}

Fig.~1E summarizes the coupling physics in momentum space for the artificially anisotropic cladding waveguide platform. The red surface represents the dispersion manifold of the cladding TE slab mode, propagating at an arbitrary in-plane angle relative to the optical axis (oriented along \emph{x}-axis). The blue surface corresponds to the TM mode guided in the silicon core; crucially, its effective index \(N_{\mathrm{core}}\) is essentially determined by the longitudinal component \(N_x\) and exhibits negligible dependence on the transverse component \(N_y\). The intersection of these two manifolds (yellow dashed line) identifies the states sharing the same \(N_x\), thereby satisfying the phase-matching condition. When this condition is met, the core TM mode couples unidirectionally into the radiative TE slab channel, leading to pronounced radiation loss. This leakage can be extinguished at a discrete set of core widths that satisfy the destructive-interference condition prescribed by Eq.~\ref{eq:bic-condition}, causing the core TM mode collapses into a BIC. These solutions are indicated by the green markers on the phase-matching line, each associated with a specific \(N_{\mathrm{core}}\). A salient feature of the SWG cladding is that the radiated TE slab waves propagate obliquely with respect to the optical axis, carrying both \(N_x\) and \(N_y\) components. Importantly, the oblique propagation angle---and more broadly, the topology of the TE continuum manifold---can be continuously reconfigured through anisotropic engineering of the SWG without perturbing guided core state. For instance, increasing the duty cycle expands the (\(N_x\), \(N_y\)) footprint of the cladding TE dispersion, which in turn shifts the intersection towards larger \(N_y\). This ability to reshape the continuum in momentum space, independent of the guided core state, provides a direct and robust lever for engineering leakage dynamics and radiation fields.

We begin with a symmetric configuration, where the optical axes of the two SWG slabs are aligned with the waveguide axis and share an identical duty cycle. This symmetry mirrors the leakage processes at both core--cladding interfaces, ensuring they are equivalent. Fig.~2A maps the calculated core TM mode transmission loss across the joint parameter space of SWG duty cycle and core width. Although the core TM mode is strongly attenuated across most of this domain, two types of low-loss regions emerge. The bottom right region corresponds to conventional total internal reflection (TIR) based waveguide. In this region, the combination of small duty cycles and large core widths ensure the existence of the guiding condition \(N_{\mathrm{TM}}>N_{\mathrm{TE}}^{\mathrm{clad}}\). Fig.~2B shows the line cut at duty cycle \(a/\Lambda=0.43\) (green dashed line in Fig.~2A). Once the core width exceeds 1200 nm, the index ordering reverses to \(N_{\mathrm{TM}}>N_{\mathrm{TE}}^{\mathrm{clad}}\), closing the radiative channel and causing the transmission loss to drop sharply before stabilizing at 0.04 dB/cm. Field profiles Figs.~2D and 2E for a representative wide-core device (\(W=1600\) nm, green triangle in Fig.~2A) confirm this formation of conventional fundamental TM mode.

Beyond the extended low-loss band of conventional guidance, Fig.~2A reveals two nearly linear discrete trajectories of loss minima that persist across a broad span of SWG duty cycles. Unlike guided modes---protected by TIR once the radiation channel is closed---these minima are intrinsically interference-enabled. They arise from the exact cancellation of radiation emitted at the two core--cladding junctions, and thus remain highly sensitive to the geometric phase accumulated within the radiative TE channel.

\begin{figure}[!t]
\centering
\includegraphics[width=\textwidth,height=0.68\textheight,keepaspectratio]{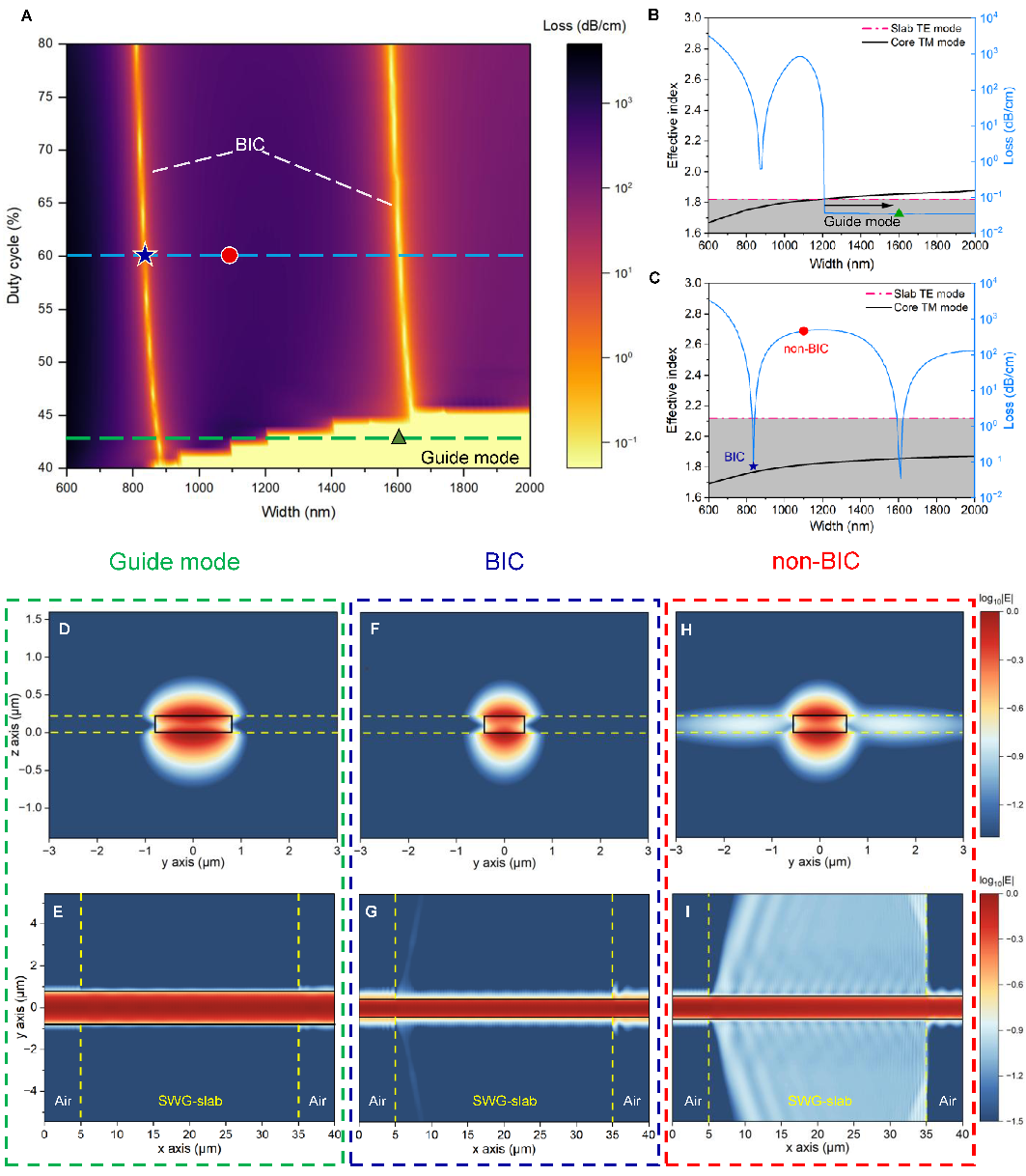}
\caption{\textbf{Simulation results of artificially anisotropic cladding waveguides.} (\textbf{A)} Transmission loss of the core TM mode in the parameter space of SWG duty cycle and core width. (\textbf{B}) Line cut corresponding to the green dashed line in (A) (duty cycle \(a/\Lambda=0.43\)). (\textbf{C}) Line cut corresponding to the blue dashed line in (A) (duty cycle \(a/\Lambda=0.6\)). (\textbf{D, F, H}) Cross-sectional field distributions of the guide mode, BIC state and non-BIC state, respectively. (\textbf{E, G, I}) Field distributions along the propagation direction for the guide mode, BIC state and non-BIC state, respectively.}
\label{fig:2}
\end{figure}

A practically significant trend emerges: as the duty cycle increases, the loci of loss minima shift systematically towards narrower core widths, demonstrating that the BIC condition can be ``steered'' through metamaterial design. Physically, increasing the duty cycle drives the SWG cladding towards a silicon-like slab response, modifying the phase-matching condition and reduces the phase difference \(\Delta\Phi\) in the interference condition governed by Eq.~\ref{eq:bic-condition}. To clarify the formation of loss minima explicitly, Fig.~2C plots the transmission loss together with the modal effective indices for a representative duty cycle of \(a/\Lambda=0.6\) (blue dashed line in Fig.~2A). Throughout the width range, the core TM mode remains embedded in the cladding TE continuum (\(N_{\mathrm{TM}}<N_{\mathrm{TE}}^{\mathrm{clad}}\)), rendering it generically leaky. Remarkably, however, the loss collapses at \(W=835\) nm and \(W=1610\) nm, reaching 0.08 dB/cm and 0.04 dB/cm, respectively---in excellent agreement with Eq.~\ref{eq:bic-condition}. Electric-field profiles (Figs.~2F--G) corroborate this, showing tight confinement with only minor leakage at the strip-to-SWG transition due to the fundamental mismatch between index-guided and interference-stabilized modes. In sharp contrast, the non-BIC state at \(W=1100\) nm (red dot in Fig.~2A) displays a pronounced radiation field within the SWG slabs (Figs.~2H--I). Here, the radiated power is carried predominantly by the in-plane field components \(E_x\) and \(E_y\), while \(E_z\) remains largely localized within the core (Supplementary Fig.~S2), consistent with leakage into an obliquely propagating TE slab channel. Because the TE wave-vector is tilted relative to the optical axis, \(E_x\) and \(E_y\) are intrinsically coupled. Consequently, the resulting BIC is inherently vectorial---necessitating the simultaneous cancellation of the entire electromagnetic field rather than a single scalar component \cite{ref20,ref32}.

\subsection{Experimental demonstration}

To validate our platform under realistic fabrication constraints, we designed and fabricated two sets of artificial anisotropic cladding waveguides with a series of duty cycles. Within each set, we systematically swept the core width to locate the geometries exhibiting a pronounced loss dip---our experimental signature of the BIC condition. Each test structure consists of a 1-mm-long waveguide section terminated by two TM grating couplers for in- and out-coupling. The device layout is schematized in Fig.~3A, with a representative scanning electron microscopy (SEM) image shown in Fig.~3B.

A practical subtlety in our design is the treatment of the cladding slabs. As highlighted by the zoomed-in SEM in Fig.~3C, lateral tapers are implemented to connect the SWG slabs to the surrounding silicon slab region. These tapers are not merely a fabrication convenience---they are essential for faithfully realizing the intended radiative continuum within a finite chip footprint. In an idealized model, the slab region is effectively unbounded, allowing radiated TE waves to propagate away at arbitrary in-plane angles and forming a true continuum of radiation states. However, if the SWG slabs are abruptly truncated, this continuum collapses into a discrete set of cladding modes, fundamentally altering the underlying physics. In such a scenario, energy is no longer irreversibly leaked through unidirectional mode conversion, but can instead recouple through hybridization between core and cladding eigenmodes (Supplementary Fig.~S3). Fig.~3D confirms this by comparing simulated loss curves for three cases: an ideal structure with effectively infinite slabs, a practical structure with lateral slab tapers, and a structure with truncated slabs. The results for the ``infinite-slab'' and ``tapered-slab'' cases nearly overlap, demonstrating that lateral tapering provides an experimentally feasible route to emulate an open radiation path. In stark contrast, truncating the slabs produces a qualitatively different response, resembling a guided-mode regime with uniformly low loss across the width sweep. In that case, the SWG region contributes to the lateral confinement of the core TM mode, suppressing radiation leakage and eliminating the interference-enabled waveguiding behavior targeted in this work.

\begin{figure}[!t]
\centering
\includegraphics[width=\textwidth,height=0.68\textheight,keepaspectratio]{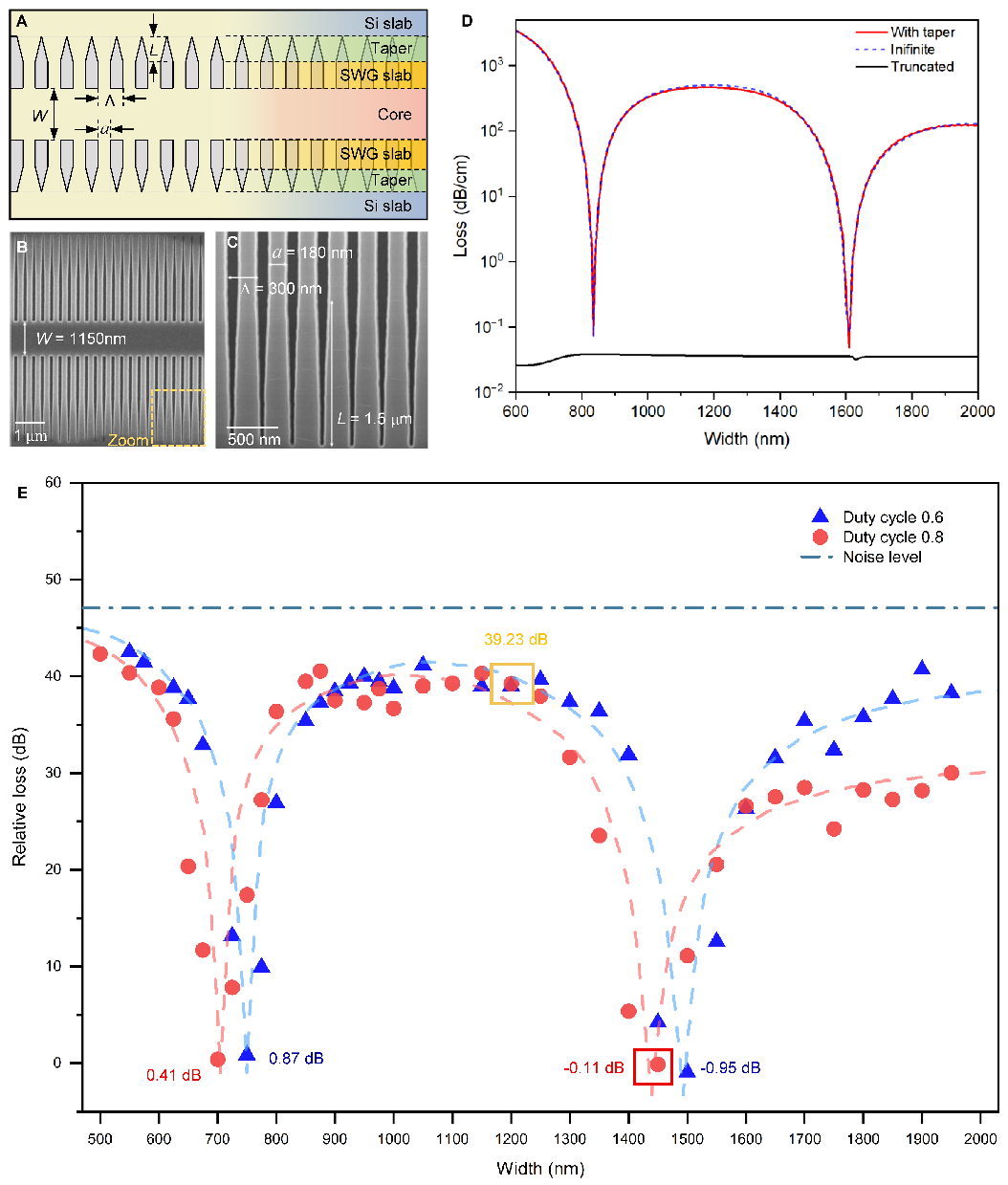}
\caption{\textbf{Experimental results of artificially anisotropic cladding waveguides.} (\textbf{A)} schematic of the practical waveguide where lateral tapers are used. (\textbf{B}) SEM of our waveguide with a duty cycle of 0.6. (\textbf{C}) Zoom-in SEM of (B). (\textbf{D}) Loss curve of core TM mode for waveguides with SWG slabs and lateral tapers, infinite SWG slabs and truncated SWG slabs. (\textbf{E}) Experimental measurement of transmission loss of core TM mode with a duty cycle of 0.6 and 0.8.}
\label{fig:3}
\end{figure}

The measured relative transmission loss for both fabricated waveguide sets is summarized in Fig.~3E. We define ``relative loss'' by comparing the power transmitted through each artificially anisotropic cladding waveguide to that of a reference strip waveguide with a matched core width, thereby factoring out the width-dependent coupling efficiency of the TM grating couplers. Markers represent data for duty cycles \(a/\Lambda=0.6\) (blue) and 0.8 (red). The overlaid curves are fits guided by the corresponding numerical simulations to emphasize the overall trend. For \(a/\Lambda=0.6\), two distinct minima are observed at \(W=750\) nm and \(W=1500\) nm, with relative-loss values of 0.87 dB and -0.95 dB, respectively. For \(a/\Lambda=0.8\), these minima occur at \(W=700\) nm and \(W=1450\) nm, yielding 0.41 dB and -0.11 dB. Note that the negative relative loss values are attributed to fabrication deviations.

Outside of these discrete widths, the transmission exhibits a sharp collapse. Specifically, in the ranges \(W\in[500,600]\) nm and \(W\in[850,1250]\) nm, the loss exceeds 35 dB, effectively reaching the noise floor of our experimental setup. This binary ``all-or-nothing'' transmission characteristic is highly consistent with the physics of interference-stabilized BICs: the mode remains strongly radiative until the specific geometry triggers destructive interference in the leakage channel. Furthermore, the systematic shift of the loss minima with duty cycle validates our central premise: that metamaterial-induced anisotropy allows for the deterministic construction of low-loss waveguides. The observed offset from theoretical predictions is likely due to the SWG period (\(\Lambda=300\) nm) being near the limit of EMT applicability. Moving to a more deeply subwavelength regime (\(\Lambda<220\) nm) would improve the predictive accuracy of the EMT model and facilitate even tighter agreement between theory and experiment.

Two distinct spectral signatures emerge from our measurements, highlighting the complementary functionalities enabled by this platform. First, the loss minima in our waveguides exhibit an intrinsically broadband low-loss response (Supplementary Fig.~S4). For instance, at a duty cycle of \(a/\Lambda=0.8\), the loss spectrum for \(W=1450\) nm (red box in Fig.~3E) remains notably flat over a wide wavelength range. This stability likely originates from the weak net dispersion of the phase condition: as the wavelength \(\lambda\) increases, the simultaneous increase in the effective-index separation, \(\Delta N=N_{\mathrm{TE}}^{\mathrm{core}}-N_{\mathrm{TM}}\), allows the destructive-interference requirement to remain approximately satisfied across a broad band. This behavior provides a direct route to broadband TM-mode transmission on platforms where TM operation is otherwise severely penalized by radiation leakage \cite{ref10}. Simultaneously, the radiative TE channel activated by phase matching is strongly dispersive, a characteristic that is advantageous rather than detrimental for resonant devices \cite{ref9,ref11}. The SWG metamaterial introduces an additional, continuously tunable anisotropic degree of freedom, enabling enhanced dispersion shaping of the continuum channel \cite{ref33,ref34}. This opens a pathway to higher-Q waveguide resonators and, correspondingly, sensors with improved spectral sharpness and sensitivity.

Second, the non-BIC regime exhibits the inverse behavior: broadband and high loss characterized by pronounced fluctuations. As shown in Fig.~3E, the spectrum recorded at \(W=1200\) nm (yellow box) is noise-like and remains strongly attenuated across the measured band. We attribute these fluctuations to unavoidable scattering and parasitic reflections encountered by the radiated TE waves as they interact with the substrate and adjacent structures. These interactions convert a nominally smooth radiative leakage into a speckle-like transmission response. Importantly, this ``always-leaky'' operating regime is itself useful: with appropriate control over the radiation field, the non-BIC region provides a robust design space for broadband polarizers \cite{ref15,ref21}. Such devices could offer a high PER over a wide bandwidth with relaxed fabrication tolerances.

\subsection{New features in anisotropy-induced asymmetric waveguides}

In conventional ridge-waveguide implementations, BIC formation typically presumes identical slab claddings on both sides of the core. Breaking the symmetry is non-trivial; it usually requires differentiating the cladding slabs geometrically---for example, by employing unequal shallow-etch depths. Such strategies offer limited flexibility in reshaping the radiative continuum and often comes at the expense of increased process complexity and more stringent fabrication tolerances.

Our artificially anisotropic cladding waveguides platform introduces a fundamentally different approach: symmetry breaking is achieved within a single etch step by simply tilting the dielectric segments by an angle \(\theta_r\). Fig.~4A illustrates this asymmetric configuration in top view, where the SWG slabs in both cladding regions are rotated in the same clockwise direction. While the two claddings remain identical in composition and can still be homogenized as the same anisotropic medium, the mirror symmetry of the optical axis relative to the waveguide core is intentionally broken. The asymmetry originates from anisotropic engineering (optical-axis control) rather than geometric non-uniformity of the claddings. This distinction is critical: optical-axis rotation introduces a continuous and highly effective control knob that expands the accessible design space for modal birefringence and radiative continuum.

This mechanism is illustrated in the momentum diagram of Fig.~4B for duty cycle \(a/\Lambda=0.6\). The effective index of cladding TE slab mode, \(N_{\mathrm{TE}}^{\mathrm{clad}}\), exhibits a strong angular dependence, reaching its maximum value of 2.140 when the mode's dominant electrical component \(E_y\) is aligned with the ordinary refractive index \(n_{\parallel}\). As the wavevector deviates from the optical axis, \(N_{\mathrm{TE}}^{\mathrm{clad}}\) decreases sharply due to the transition of the influencing index from \(n_{\parallel}\) to \(n_{\perp}\). \(N_{\mathrm{TE}}^{\mathrm{clad}}\) reaches its minimum value of 1.506 when the TE wave transmission is perpendicular to the optical axis. In stark contrast, the effective index of the core TM mode remains essentially invariant (\(N_{\mathrm{TM}}=1.802\)), as the corresponding index profile of \(E_z\) is unaffected by rotations of optical axis within the \emph{x}-\emph{y} plane. Consequently, the TE continuum can be swept across the nearly fixed TM dispersion, leading to a critical tilt angle \(\theta_{\mathrm{lim}}=42^\circ\) beyond which \(N_{\mathrm{TE}}^{\mathrm{clad}}<N_{\mathrm{TM}}\) and the leaky TM state transitions into a guided regime. The same physics persists---and becomes even more tunable---at higher duty cycle. For \(a/\Lambda=0.8\), both \(n_{\parallel}\) and \(n_{\perp}\) increase, expanding the attainable index range and the usable rotation window. As shown in Fig.~4C, \(N_{\mathrm{TE}}^{\mathrm{clad}}\) spans from 2.496 to 1.618, while \(N_{\mathrm{TM}}\) remains essentially constant at 1.807. Correspondingly, the critical tilt angle increases to \(\theta_{\mathrm{lim}}=64^\circ\). Together, these results establish optical-axis rotation in SWG slabs as a practical, fabrication-light, and highly leveragable degree of freedom for engineering asymmetric continua that are difficult to realize in conventional etched-slab platforms.

To explore the asymmetric regime under realistic fabrication constraints, we focus on devices with duty cycle \(a/\Lambda=0.6\), which provides robust pattern fidelity while retaining strong anisotropic tunability. Fig.~4D maps the calculated transmission loss as a function of core width and tilt angle \(\theta_r\). Two trends are immediately evident. First, the discrete, geometry-selected loss minima persist over a broad range of \(\theta_r\), indicating that the interference-enabled confinement mechanism remains operative even after symmetry breaking. Second, within this interference-based low-loss region, the minimum loss increases with tilt angle, signaling that anisotropy-induced asymmetry progressively weakens the confinement of the bound state supported under the symmetric conditions and couples it back to the radiation continuum. Once the SWG segments are tilted, the leakage processes at the two core--cladding junctions are no longer equivalent. In particular, the radiative TE slab waves emitted into the left and right claddings propagate with different orientations relative to the local optical axis, and therefore obey different phase-matching conditions. The corresponding radiation channels are characterized by distinct effective indices \(N_{\mathrm{TE}}^{\mathrm{clad}}\) and emission angles \(\theta\); crucially, the transmitted and reflected TE components generated at each junction acquire different amplitudes and phases \cite{ref31}. Consequently, the destructive-interference condition can no longer be satisfied simultaneously on both sides, resulting in incomplete radiation cancellation and transforming the lossless guiding state into a finite-loss regime with tunable leakage.
\begin{figure}[!t]
\centering
\includegraphics[width=\textwidth,height=0.68\textheight,keepaspectratio]{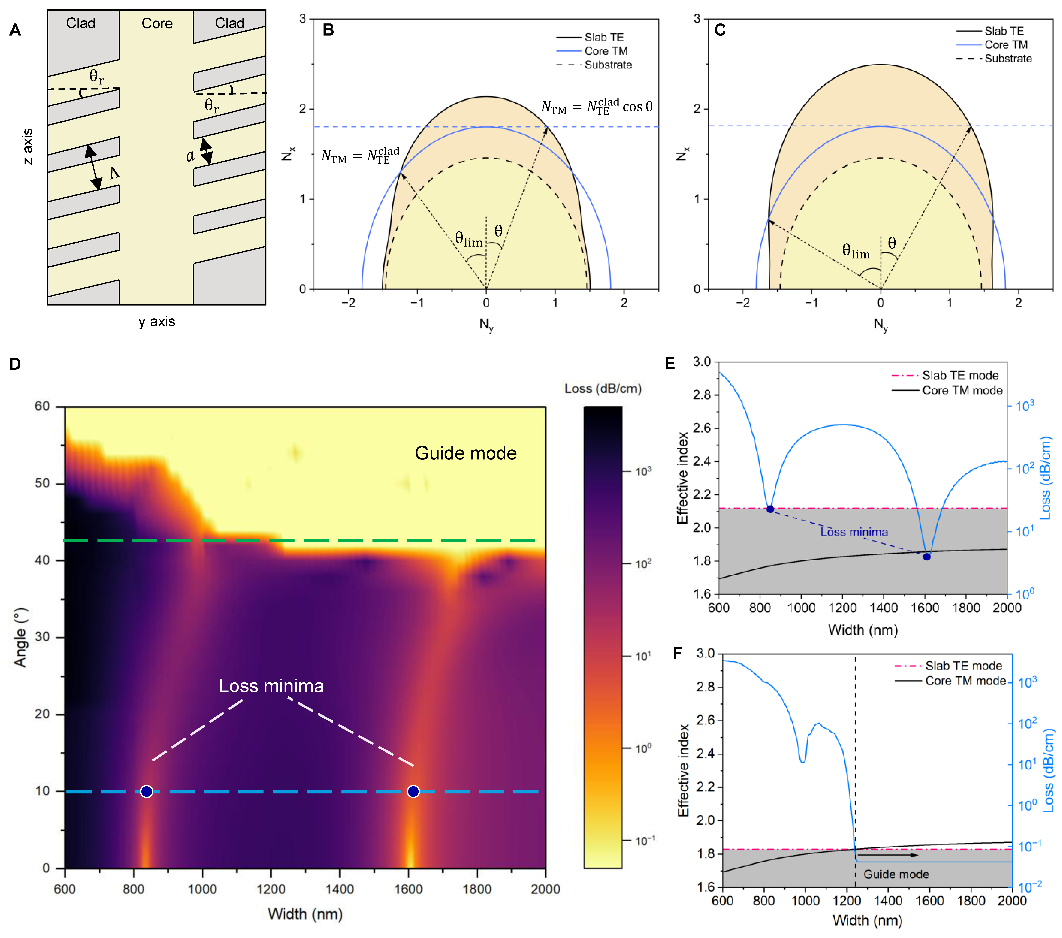}
\caption{\textbf{Simulation results of anisotropy-induced asymmetric waveguides.} (\textbf{A)} Schematic of the asymmetric waveguide where the segments of the SWG rotates along the same clockwise at a tilt angle \(\theta_r\). (\textbf{B}) momentum of modes in asymmetric waveguides with a duty cycle of 0.6. (\textbf{C}) momentum of modes in asymmetric waveguides with a duty cycle of 0.8. (\textbf{D}) Transmission loss of core TM modes in the parameter space of tilt angle and core width for asymmetric waveguides with a duty cycle of 0.6. Line cut of loss curve and modes' effective indices for a tilt angle (\textbf{E}) \(10^\circ\) and (\textbf{F}) \(25^\circ\).}
\label{fig:4}
\end{figure}

Fig.~4E shows a representative line cut at \(\theta_r=10^\circ\), plotted alongside the corresponding effective-index evolution. Two distinct loss minima remain visible at \(W=850\) nm and \(W=1610\) nm; however, their values rise to 22 dB/cm and 4 dB/cm, respectively---orders of magnitude higher than those observed in the symmetric configuration. Crucially, the condition \(N_{\mathrm{TE}}^{\mathrm{clad}}>N_{\mathrm{TM}}\) is maintained across the entire width sweep, confirming that the radiation channel remains open. This indicates that the observed minima originate from imperfect destructive interference rather than a fundamental closure of the continuum. Consistent with this interpretation, the radiation-field strengths in the two cladding slabs become markedly unequal regime (Supplementary Fig.~S5), evidencing that the left and right junctions radiate with different amplitudes and phases.

Interestingly, beyond a certain tilt angle, the minimum loss begins to decrease. As the tilt angle approaching \(\theta_{\mathrm{lim}}\), the reduced value of \(N_{\mathrm{TE}}^{\mathrm{clad}}\) causes the TE waves to radiate at a smaller exit angle, nearing a glancing incidence. This results in an amplitude reduction of the coupled components and weakens the TM-to-TE conversion efficiency \cite{ref2,ref31}, thereby reducing the strength of the radiative leakage even when interference remains imperfect. Ultimately, at sufficiently large tilt angles, \(N_{\mathrm{TE}}^{\mathrm{clad}}\) is lowered enough that the radiation channel closes (\(N_{\mathrm{TM}}>N_{\mathrm{TE}}^{\mathrm{clad}}\)), and the leaky state transitions into a conventional guided regime. This evolution is illustrated in Fig.~4F for \(\theta_r=42^\circ\) (green dashed line in Fig.~4D). An interference-derived loss minimum persists near \(W=960\) nm with a loss of 10 dB/cm, while the continuum remains open (\(N_{\mathrm{TE}}^{\mathrm{clad}}>N_{\mathrm{TM}}\)). Once the core width exceeds 1240 nm, the dispersion ordering reverses to \(N_{\mathrm{TM}}>N_{\mathrm{TE}}^{\mathrm{clad}}\), the radiative channel is cut off, and the transmission loss drops sharply at the transition width as the mode becomes guided.

To experimentally probe the new regimes enabled by anisotropy-induced symmetry breaking, we fabricated two sets of anisotropy-induced asymmetric waveguides with tilt angles of \(\theta_r=10^\circ\) and \(25^\circ\). Within each set, the core width was systematically varied while the SWG duty cycle was held nominally constant, enabling a direct assessment of how tilt-driven asymmetry reshapes the loss landscape. Fig.~5A illustrates the schematic of these fabricated devices, where the duty cycle is defined along the grating direction and a complementary tilt angle is adopted for metrological convenience. Lateral tapers were also integrated to ensure unencumbered radiation into slab regions. The SEM of the asymmetric waveguides with a tilt angle \(\theta_r=10^\circ\) is shown in the Fig.~5B as an example, where the width of the core is measured to be 800 nm and symmetry of the waveguide is successfully broken by tilted SWGs whose optical axis are parallel to each other. The zoom-in graph of tilted SWG-slab is provided in the Fig.~5C. Along the grating direction, the SWG has a period of 300 nm and width of silicon segment is 180 nm, corresponding to a duty cycle \(a/\Lambda=0.6\). Similarly, for \(\theta_r=25^\circ\), the SWG slabs feature a period of 270 nm and a segment width of 160 nm (provided in Supplementary Fig.~S6).

\begin{figure}[!t]
\centering
\includegraphics[width=\textwidth,height=0.68\textheight,keepaspectratio]{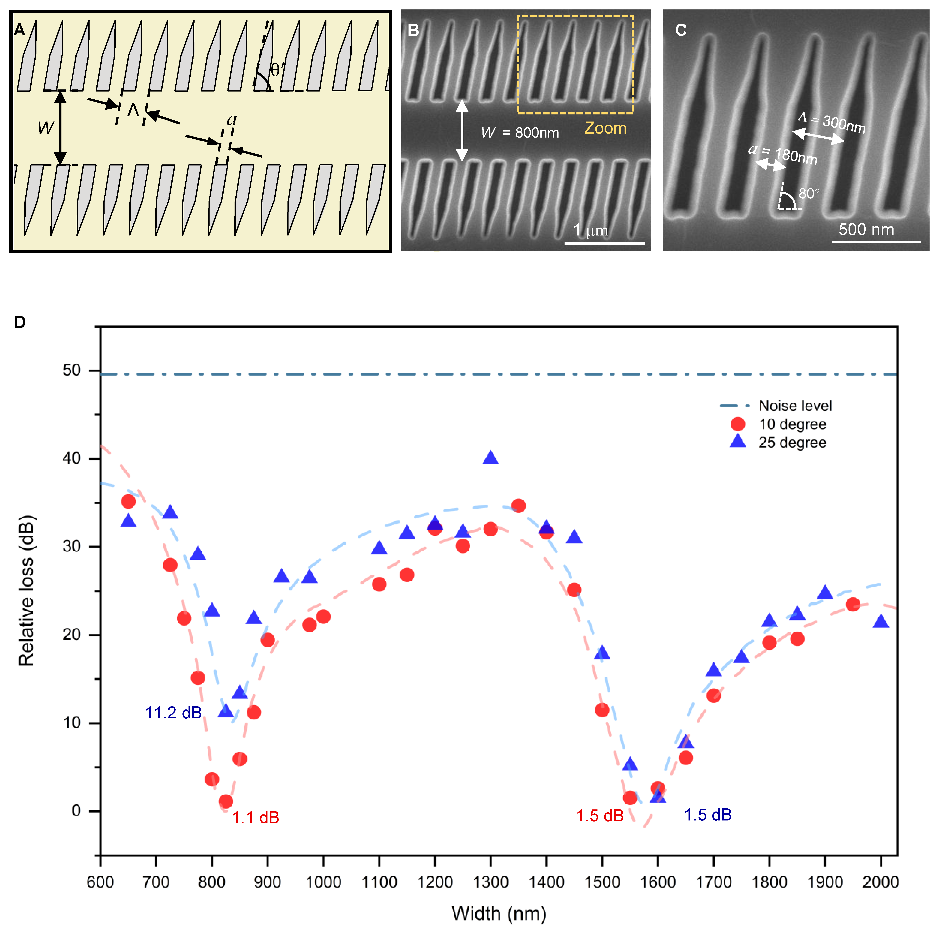}
\caption{\textbf{Experimental results of anisotropy-induced asymmetric waveguides.} (\textbf{A)} schematic of the practical asymmetric waveguide, where the lateral tapers are used to ensure unencumbered radiation and a complementary tilt angle is adopted to metrological convenience. (\textbf{B}) SEM of the practical asymmetric waveguide with a duty cycle of 0.6 and tilt angle \(10^\circ\). (\textbf{C}) Zoom-in SEM of (B). (\textbf{D}) Experimental measurement of transmission loss of core TM mode with a tilt angle of \(10^\circ\) and \(25^\circ\).}
\label{fig:5}
\end{figure}

Fig.~5D summarizes the measured relative transmission loss for the asymmetric devices. For the \(\theta_r=10^\circ\) set, two pronounced minima are observed at \(W=825\) nm and \(W=1550\) nm, with relative- loss values of 1.1 dB and 1.5 dB, respectively. Increasing the tilt angle to \(\theta_r=25^\circ\) preserves the existence of discrete minima but markedly reshapes their depths, with loss dips of 11.2 dB and 1.5 dB occurring at \(W=825\) nm and \(W=1600\) nm. Simulation-guided fits (dashed curves) are overlaid to highlight the overall evolution of the loss landscape as asymmetry increases. The observed loss minima in asymmetric waveguides confirm the robustness of the interference-enabled guiding regime. The distinct positions and magnitudes of the loss-minimum associated with different tilt angles further demonstrate that anisotropy-induced asymmetry introduces a powerful degree of freedom for controlling the destructive interference of radiative waves and the overall radiation strength. Distinct from conventional ridge waveguides, the radiation-controllable waveguides demonstrated here are realized through anisotropic metamaterial engineering in the cladding, rather than geometric displacement. The additional control enables precise tailoring of the radiative continuum, giving rise to diverse leakage characteristics that can be adapted to specific application requirements. Moreover, such anisotropy-induced asymmetry offers a compelling platform to explore emergent physical phenomena in cladding--core--cladding waveguides, including unidirectional resonances \cite{ref35,ref36,ref37} and asymmetric BICs \cite{ref38,ref39,ref40}, which have been reported in other photonic platforms.

In conclusion, our scheme enables deterministic and versatile construction of BIC-induced waveguiding by integrating SWG slabs into the cladding region. Unlike conventional ridge structures, where the radiative continuum is formed through a vertical one-dimensional design, our approach utilizes a lateral two-dimensional design, significantly broadening the design freedom. The strong and tunable anisotropy of the metamaterial serves as a versatile tool to engineer both the radiative continuum and leakage behavior without introducing additional fabrication complexity. This capability holds great potential for advanced photonic applications, such as ultra-sensitive sensors, high PER polarizers, and broadband TM-operation devices. Furthermore, the anisotropy-induced asymmetry, achieved by tilting the SWG segments, offers a practical route to forming robust interference-mediated guiding regime with controllable radiation fields. By precisely engineering anisotropic asymmetry, it is promising to realize physical phenomena in the paradigm of cladding-core-cladding waveguides, such as asymmetric BICs and unidirectional resonances. Our work establishes anisotropy-induced BIC waveguiding as a practical and general design strategy that is compatible with standard lithographic patterning across diverse integrated photonic platforms, while opening new opportunities for further tailoring cladding anisotropy through advanced metamaterial designs.

\section{Materials and Methods}

\subsection{Fabrication}

Samples were fabricated on a standard SOI platform with a 220-nm silicon device layer and a 2-\(\mu\mathrm{m}\) buried oxide. Following piranha cleaning, a 300-nm-thick diluted positive resist (ZEP520A) was spin-coated and baked at \(180^\circ\mathrm{C}\) for 5 min. Grating couplers and our waveguides were defined by electron-beam lithography (NanoBeam, nB5). After exposure, the resist was developed in amyl acetate for 100 s and rinsed in isopropyl alcohol for 30 s. Pattern transfer into the silicon layer was performed using inductively coupled plasma (ICP) etching with an etch selectivity of \(\sim\)1:1 relative to the resist, followed by resist removal in dimethyl sulfoxide to obtain the final structures.

\subsection{Characterization}

The fabricated symmetric and asymmetric waveguides were characterized using scanning electron microscopy (SEM). Symmetric structures were imaged using a HITACHI SU8010 SEM, confirming identical duty cycles and optical-axis orientations of the SWGs in both cladding regions, with the optical axis aligned parallel to the core--cladding interface. Asymmetric waveguides were characterized using an Apreo 2S HiVac SEM, where the SWG segments were uniformly tilted in the clockwise direction. Optical transmission measurements were performed using a tunable continuous-wave laser (Santec TSL-550) with a wavelength step of 5 pm over the 1500--1600 nm range. The transmitted power was recorded using a power meter (Santec MPM-210H) with a detection limit of -70 dBm. Input and output fibers were aligned using a six-degree-of-freedom positioning stage to ensure optimal coupling. Owing to the broadband low-loss response of loss minima and the strongly fluctuating spectra associated with non-BICs (see Supplementary Fig.~S4), the average transmitted power within a 3-dB bandwidth of the strip waveguides was used to quantify the relative propagation loss for each measurement.

\section*{Acknowledgments}
We thank Prof. Lluis Torner and Prof. David Artigas (Universitat Polit\`{e}cnica de Catalunya, Spain) for valuable discussions. We also thank the ShanghaiTech Material Device Lab (SMDL) for technical support.

\section*{Funding}
This work was supported by the National Natural Science Foundation of China (NSFC) (U22A2093); Mobile Information Networks-National Science and Technology Major Project (2025ZD1302900); Science, Technology and Innovation Commission of Shenzhen Municipality (JCYJ20210324131614040, GXWD20231130113557001, KJZD20240903100009013, KJZD20240903101100002); National Key Laboratory of Laser Spatial Information Foundation (LSI2024JCKY02); and Key Research and Development Program of Ningxia Hui Autonomous Region (2025BEG01003).

\section*{Author Contributions}
X.X., Y.Z., and Y.D. conceived the idea; J.W., K.L., and Y.L. performed the theoretical analysis and numerical modeling; X.X., J.W., K.L., W.Y., Y.F., and Y.C. designed the devices and conducted the experiments; X.X., J.W., and K.L. analysed the data; X.X., Y.Z., Y.D., W.L., F.H., and J.D. supervised the project; X.X., J.W., and K.L. wrote the manuscript; All authors contributed to the discussion of the results and writing of the manuscript.

\section*{Competing Interests}
The authors declare that they have no competing interests.

\section*{Data and Materials Availability}
All data needed to evaluate the conclusions in the paper are present in the paper and/or the Supplementary Materials.

\end{document}